\documentclass{JHEP3}

\title{String background fields and Riemann-Cartan geometry}
\author{Milovan Vasili\'c \\
        Institute of Physics, P.O.Box 57, 11001 Belgrade, Serbia \\
        E-mail: \email{mvasilic@ipb.ac.rs}}

\abstract{ We study classical dynamics of cylindrical membranes wrapped
around the extra compact dimension of a $(D+1)$-dimensional Riemann-Cartan
spacetime. The world-sheet equations and boundary conditions are obtained
from the universally valid conservation equations of the stress-energy and
spin tensors. Specifically, we consider membranes made of macroscopic matter
with maximally symmetric distribution of spin. In the narrow membrane limit,
the dimensionally reduced theory is obtained. It describes how effective
strings couple to the effective $D$-dimensional geometry. The striking
coincidence with the string theory $\sigma$-model is observed. In this
correspondence, the string background fields $G_{\mu\nu}$, $B_{\mu\nu}$,
$A_{\mu}$ and $\Phi$ are related to the metric and torsion of the
Riemann-Cartan spacetime. }


\preprint{}

\usepackage{amsmath}

\newcommand{\itGamma}{{\mathit{\Gamma}}}
\newcommand{\del}{\partial}

\newcommand{\cK}{{\cal K}}
\newcommand{\cM}{{\cal M}}

\newcommand{\cR}{{\cal R}}

\newcommand{\cT}{{\cal T}}

\newcommand{\lc}{\varepsilon}

\newcommand{\diag}{\mathop{\rm diag}\nolimits}
\newcommand{\orto}{{\scriptscriptstyle\perp}}

\newcommand{\Pn}{ {P_{\orto}}\vphantom{P} }
\newcommand{\cfl}[2]{{\textstyle {{#1}\brace {#2}}}}

\newcommand{\rmd}{d}

\begin{document}

\section{Introduction}\label{Sec1}

The problem of motion of extended objects in curved backgrounds is usually
addressed by using the multipole formalism originally proposed in
\cite{1,2}. In \cite{11,12,12a}, a manifestly covariant version of this
method has been developed, and applied to strings and higher branes in
Riemann-Cartan geometry. One starts with the covariant conservation law of
the stress-energy and spin tensors of matter fields, and analyzes it under
the assumption that matter is localized to resemble a brane. In the lowest,
single-pole approximation, the moving matter is viewed as an infinitely thin
brane. In the pole-dipole approximation, its non-zero thickness is taken
into account.

The known results concerning higher branes in Riemann-Cartan geometry can be
summarized as follows. In the spinless case, the background curvature is
coupled to the internal orbital angular momentum of a thick brane. The
coupling is shown to disappear in the limit of an infinitely thin brane
\cite{11,12}. The dynamics of spinning branes has been investigated in
\cite{12a}. The spin-torsion coupling has been derived for an arbitrary
distribution of spinning matter over the brane. In \cite{12b}, the general
results of \cite{12a} have been applied to infinitely thin membranes.
Specifically, the membranes with maximally symmetric distribution of
stress-energy and spin have been considered. In the spinless case, such
membranes are characterized by the tension alone, and are known as the
Nambu-Goto membranes \cite{13,14}. These membranes do not couple to torsion,
but their minimal extension characterized by a uniformly distributed spin,
does. It has been shown in \cite{12b} that these membranes couple to torsion
the same way as string theory membranes couple to the string theory 3-form
field \cite{22}. Strings have been introduced by considering cylindrical
membranes wrapped around the extra compact dimension of a
$(D+1)$-dimensional spacetime. After a $D+1\to D$ dimensional reduction, the
effective string dynamics has been obtained. The corresponding action
functional has been verified to share many features with the string theory
$\sigma$-model action of \cite{15,16,17,18,t7,t19}. In particular, the
effective string couples to spacetime metric and torsion the same way as
fundamental strings couple to the string background fields $G_{\mu\nu}$ and
$B_{\mu\nu}$.

In this paper, we continue investigating the behavior of membranelike
extended objects in Riemann-Cartan spacetime. Our effort is motivated by the
partial success of \cite{12b} to relate spacetime metric and torsion to the
string background fields. In \cite{12b}, this has been achieved by viewing
strings as dimensionally reduced cylindrical membranes made of uniformly
distributed spinning matter. Precisely, the distribution of the membrane
stress-energy and spin has been attributed maximal symmetry. As a
consequence, the dimensional reduction turns them into strings characterized
by two constants only: the tension and spin magnitude. Such strings are
minimally extended Nambu-Goto strings, which turn out to couple to spacetime
metric and torsion the same way as fundamental strings couple to the string
background fields $G_{\mu\nu}$ and $B_{\mu\nu}$.

In what follows, we shall continue the analysis of cylindrical membranes by
relaxing the condition of maximal symmetry used in \cite{12b}. The reason
for this is twofold. First, what we are really interested in is the
effective string obtained in the narrow membrane limit, not the membrane
itself. This means that the membrane constituent matter does not have to be
uniformly distributed along the membrane compact dimension. Indeed, the
violation of maximal symmetry in this direction disappears after dimensional
reduction. The effective string can still be maximally symmetric, possibly
determined with more than two free parameters. The second reason is that the
effective strings characterized by additional parameters may have new
couplings with the dimensionally reduced background. In fact, it is well
known that, beside the $D$-dimensional metric, the $D+1\to D$ dimensional
reduction leaves us with additional vector and scalar fields. It is our hope
that these fields can be related to the string background fields $A_{\mu}$
and $\Phi$ very much the same as metric and torsion of \cite{12b} have been
related to $G_{\mu\nu}$ and $B_{\mu\nu}$.

We emphasize that our work is not a part of the mainstream string theory
considerations. Our membranelike objects are made of conventional matter,
and are used to probe Riemann-Cartan geometry. The only connection with
string theory is seen in the form of our resulting equations. Namely, {\it
we demonstrate in this paper that our macroscopic string couples to the
spacetime metric and torsion the same way as fundamental string couples to
the low-energy string fields} \cite{15,16,17,18,t7,t19}. In this
correspondence, the string background fields $G_{\mu\nu}$, $A_{\mu}$ and
$\Phi$ are related to the spacetime metric, while the string axion
$B_{\mu\nu}$ is related to torsion. Whether this is just a coincidence, or
there is more content in this analogy is not the subject of this paper.

The layout of the paper is as follows. In section \ref{Sec2}, we review the
basic notions of the multipole formalism developed in \cite{11,12,12a}. The
manifestly covariant world-sheet equations and boundary conditions are
explicitly displayed. The brane stress-energy and spin tensors appear in the
world-sheet equations as free coefficients. In section \ref{Sec3}, the
general equations are applied to the membrane case. The membrane spin tensor
is chosen maximally symmetric, whereas its stress-energy is left arbitrary.
In section \ref{Sec4}, a narrow membrane wrapped around the extra compact
dimension of a $(D+1)$-dimensional spacetime is considered. The membrane
stress-energy is chosen maximally symmetric only in large dimensions, while
left arbitrary in the compact dimension. In section \ref{Sec5}, the obtained
string dynamics is compared with the string $\sigma$-model of
\cite{15,16,17,18,t7,t19}. Section \ref{Sec6} is devoted to concluding
remarks.

Our conventions are the same as in \cite{12}. Greek indices $\mu,\nu,\dots$
are the spacetime indices, and run over $0,1,\dots,D-1$. Latin indices
$a,b,\dots$ are the world-sheet indices and run over $0,1,\dots,p$. Latin
indices $i,j,\dots$ refer to the world-sheet boundary and take values
$0,1,\dots,p-1$. The coordinates of spacetime, world-sheet and world-sheet
boundary are denoted $x^{\mu}$, $\xi^a$ and $\lambda^i$ respectively. The
spacetime metric is denoted by $g_{\mu\nu}(x)$, and the induced world-sheet
metric by $\gamma_{ab}(\xi)$. The signature convention is defined by
$\diag(-,+,\dots,+)$, and the indices are raised using the inverse metrics
$g^{\mu\nu}$ and $\gamma^{ab}$.

\section{Multipole formalism}\label{Sec2}

It has been shown in \cite{11,12} that an exponentially decreasing function
can be expanded as a series of $\delta$-function derivatives. For example, a
tensor-valued function $F^{\mu\nu}(x)$, well localized around the
$p+1$-dimensional surface $\cM$ in $D$-dimensional spacetime, can be
decomposed in a manifestly covariant way as
\begin{eqnarray}
F^{\mu\nu}(x) & = & \int_{\cM} d^{p+1}\xi \sqrt{-\gamma} \left[
M^{\mu\nu}\frac{\delta^{(D)}(x-z)}{\sqrt{-g}}\right. \nonumber \\ &&
\left.-\nabla_{\rho}\left( M^{\mu\nu\rho}
\frac{\delta^{(D)}(x-z)}{\sqrt{-g}} \right)+\cdots\right].      \label{jna1}
\end{eqnarray}
The surface $\cM$ is defined by the equation $x^{\mu}=z^{\mu}(\xi)$, where
$\xi^a$ are the surface coordinates, and the coefficients $M^{\mu\nu}(\xi)$,
$M^{\mu\nu\rho}(\xi), \dots$ are spacetime tensors called multipole
coefficients. Here, and throughout the paper, we make use of the surface
coordinate vectors
$$
u_a^{\mu} \equiv \frac{\del z^{\mu}}{\del\xi^a},
$$
and the surface induced metric tensor
$$
\gamma_{ab} = g_{\mu\nu} u_a^{\mu} u_b^{\nu}.
$$
The induced metric is assumed to be nondegenerate, $\gamma \equiv
\det(\gamma_{ab}) \neq 0$, and of Minkowski signature. The same holds for
the target space metric $g_{\mu\nu}(x)$ and its determinant $g(x)$. The
covariant derivative $\nabla_{\rho}$ is defined by the Levi-Civita
connection.

The decomposition (\ref{jna1}) is particularly useful for the description of
brane-like objects in spacetimes of general geometry. In \cite{12a}, it has
been used to model the fundamental matter currents: stress-energy
$\tau^{\mu}{}_{\nu}$, and spin tensor $\sigma^{\lambda}{}_{\mu\nu}$. The
brane dynamics in Riemann-Cartan backgrounds is obtained from the covariant
conservation equations \cite{23}
\begin{subequations} \label{jna2}
\begin{equation}\label{jna2a}
\left( D_{\nu} + \cT^{\lambda}{}_{\nu\lambda} \right)
\tau^{\nu}{}_{\mu} = \tau^{\nu}{}_{\rho}\cT^{\rho}{}_{\mu\nu} +
\frac{1}{2}\sigma^{\nu\rho\sigma}\cR_{\rho\sigma\mu\nu},
\end{equation}
\begin{equation}\label{jna2b}
\left( D_{\nu} + \cT^{\lambda}{}_{\nu\lambda} \right)
\sigma^{\nu}{}_{\rho\sigma} =
\tau_{\rho\sigma}-\tau_{\sigma\rho}.
\end{equation}
\end{subequations}
Here, $D_{\nu}$ is the covariant derivative with the nonsymmetric connection
$\itGamma^{\lambda}{}_{\mu\nu}$, which acts on a vector $v^{\mu}$ according
to the rule $D_{\nu}v^{\mu} \equiv \del_{\nu}v^{\mu} +
\itGamma^{\mu}{}_{\lambda\nu}v^{\lambda}$. The torsion
$\cT^{\lambda}{}_{\mu\nu}$ and curvature $\cR^{\mu}{}_{\nu\rho\sigma}$ are
defined in the standard way:
\begin{eqnarray}
\cT^{\lambda}{}_{\mu\nu} & \equiv & \itGamma^{\lambda}{}_{\nu\mu} -
\itGamma^{\lambda}{}_{\mu\nu},                                \nonumber \\
\cR^{\mu}{}_{\nu\rho\sigma} & \equiv &
\del_{\rho}\itGamma^{\mu}{}_{\nu\sigma} \! -
\del_{\sigma}\itGamma^{\mu}{}_{\nu\rho} +
\itGamma^{\mu}{}_{\lambda\rho}\itGamma^{\lambda}{}_{\nu\sigma} \! -
\itGamma^{\mu}{}_{\lambda\sigma}\itGamma^{\lambda}{}_{\nu\rho} . \nonumber
\end{eqnarray}
The covariant derivative $D_{\nu}$ is assumed to satisfy the metricity
condition $D_{\lambda}g_{\mu\nu}=0$. As a consequence, the connection
$\itGamma^{\lambda}{}_{\mu\nu}$ is split into the Levi-Civita connection and
the contortion:
\begin{eqnarray}
\itGamma^{\lambda}{}_{\mu\nu} & = &
\cfl{\lambda}{\mu\nu} + K^{\lambda}{}_{\mu\nu},              \nonumber \\
K^{\lambda}{}_{\mu\nu} & \equiv & -\frac{1}{2} \left(
\cT^{\lambda}{}_{\mu\nu} - \cT_{\nu}{}^{\lambda}{}_{\mu}
+ \cT_{\mu\nu}{}^{\lambda} \right) .                         \nonumber
\end{eqnarray}
The curvature tensor can then be rewritten in terms of the Riemann curvature
$R^{\mu}{}_{\nu\rho\sigma} \equiv
\cR^{\mu}{}_{\nu\rho\sigma}(\itGamma\to\{\})$, and the Riemannian covariant
derivative $\nabla_{\mu} \equiv D_{\mu}(\itGamma\to\{\})$:
$$
\cR^{\mu}{}_{\nu\lambda\rho} = R^{\mu}{}_{\nu\lambda\rho} + 2
\nabla_{[\lambda}K^{\mu}{}_{\nu\rho]} + 2 K^{\mu}{}_{\sigma[\lambda}
K^{\sigma}{}_{\nu\rho]}.
$$

Given the system of conservation equations (\ref{jna2}), one finds that the
second one has no dynamical content. One can use (\ref{jna2b}) to eliminate
$\tau^{[\mu\nu]}$ from (\ref{jna2a}), and thus obtain
\begin{equation}\label{jna3}
\nabla_{\nu} \left( \theta^{\mu\nu} - {\cal D}^{\mu\nu} \right) =
\frac{1}{2} \sigma_{\nu\rho\lambda}\nabla^{\mu} K^{\rho\lambda\nu},
\end{equation}
where $\theta^{\mu\nu}$ stands for the generalized Belinfante tensor
$$
\theta^{\mu\nu} \equiv \tau^{(\mu\nu)} -
\nabla_{\rho}\sigma^{(\mu\nu)\rho} -
\frac{1}{2}K_{\lambda\rho}{}^{(\mu}\sigma^{\nu)\rho\lambda},
$$
and ${\cal D}^{\mu\nu} \equiv K^{[\mu}{}_{\lambda\rho}
\sigma^{\rho\lambda\nu]} + \frac{1}{2} K_{\lambda\rho}{}^{[\mu}
\sigma^{\nu]\rho\lambda}$. The conservation law in the form (\ref{jna3}) is
the starting point for the derivation of brane world-sheet equations. In the
particle case, this form of the conservation equations has been used in
\cite{24}.

In this paper, we are interested in infinitely thin branes, and therefore,
restrict our analysis to the single-pole approximation. The multipole
expansion of our basic variables then reads:
\begin{subequations} \label{jna5}
\begin{equation} \label{jna5a}
\theta^{\mu\nu} = \int_{\cM} \rmd^{p+1}\xi \sqrt{-\gamma}\, T^{\mu\nu}
\frac{\delta^{(D)}(x-z)}{\sqrt{-g}} ,
\end{equation}
\begin{equation} \label{jna5b}
\sigma^{\lambda\mu\nu} = \int_{\cM} \rmd^{p+1}\xi \sqrt{-\gamma}\,
S^{\lambda\mu\nu} \frac{\delta^{(D)}(x-z)}{\sqrt{-g}} ,
\end{equation}
\end{subequations}
where $T^{\mu\nu}(\xi)$ and $S^{\lambda\mu\nu}(\xi)$ are the corresponding
multipole coefficients. The de\-com\-po\-si\-ti\-on (\ref{jna5}) is used as
an ansatz for solving the conservation equations (\ref{jna3}). This has
already been done in \cite{12a}, resulting in manifestly covariant $p$-brane
world-sheet equations.

\section{Membrane dynamics}\label{Sec3}

In \cite{12b}, the general result of \cite{12a} has been applied to the
membrane case. In particular, the membrane with maximally symmetric
distribution of spin has been thoroughly examined. In what follows, the
world-sheet equations and boundary conditions of \cite{12a} will be used as
the starting point of our analysis.

Let us start with the brief recapitulation of the known results. The
$p$-brane world-sheet equations in the single-pole approximation are
obtained in the following way. We insert the ansatz (\ref{jna5}) into the
conservation equations (\ref{jna3}), and solve for the unknown variables
$z^{\mu}(\xi)$, $T^{\mu\nu}(\xi)$ and $S^{\lambda\mu\nu}(\xi)$. The
algorithm for solving this type of equation has been discussed in detail in
\cite{12,12a}, and here we use the ready-made result. According to
\cite{12a}, the single-pole world-sheet equations are given by
\begin{subequations} \label{jna6}
\begin{equation} \label{jna6a}
\Pn^{\mu}_{\rho} \Pn^{\nu}_{\sigma} D^{\rho\sigma} =0,
\end{equation}
\begin{equation}\label{jna6b}
\nabla_a\left( t^{ab}u_b^{\mu}-2 u^a_{\rho} D^{\mu\rho} +
u_b^{\mu}u^b_{\rho} u^a_{\sigma}D^{\rho\sigma}\right) =
\frac{1}{2}S_{\nu\rho\sigma} \nabla^{\mu} K^{\rho\sigma\nu} ,
\end{equation}
\end{subequations}
while the boundary conditions have the form
\begin{equation} \label{jna7}
n_a \left( t^{ab}u_b^{\mu} - 2 u^a_{\rho} D^{\mu\rho} + u_b^{\mu} u^b_{\rho}
u^a_{\sigma} D^{\rho\sigma} \right) \Big|_{\del\cM}=0.
\end{equation}
Here, $\Pn^{\mu}_{\nu} \equiv \delta^{\mu}_{\nu} - u_a^{\mu}u^a_{\nu}$ is
the orthogonal world-sheet projector, $n^a$ is the unit boundary normal, and
$\nabla_a$ stands for the total covariant derivative that acts on both types
of indices:
$$
\nabla_a V^{\mu b} \equiv  \del_a V^{\mu b} + \cfl{\mu}{\lambda\rho}
u_a^{\rho} V^{\lambda b} + \cfl{b}{ca}V^{\mu c}.
$$
The coefficients $t^{ab}(\xi)$ and $S^{\lambda\mu\nu}(\xi)$ are the residual
free parameters of the theory. While $t^{ab}$ represents the effective
stress-energy tensor of the brane, the $S^{\lambda\mu\nu}$ currents are
related to its spin density. The shorthand notation
$$
D^{\mu\nu} \equiv K^{[\mu}{}_{\lambda\rho} S^{\rho\lambda\nu]} + \frac{1}{2}
K_{\lambda\rho}{}^{[\mu} S^{\nu]\rho\lambda}
$$
is introduced for convenience.

The world-sheet equations (\ref{jna6}) and boundary conditions (\ref{jna7})
describe the dynamics of an infinitely thin $p$-brane in $D$-dimensional
spacetime with curvature and torsion. By inspecting their form, we realize
that only branes made of spinning matter can probe spacetime torsion.
Moreover, if the spin tensor $S^{\lambda\mu\nu}$ is totally antisymmetric,
only axial component of torsion survives in the brane equations. In this
paper, we are interested in membranes characterized by maximally symmetric
distribution of spin. It has already been shown in \cite{12b} that such
membranes must have axial spin tensor of the form
\begin{equation} \label{jna8}
S^{\lambda\mu\nu} = s\,e^{abc} u_a^{\lambda} u_b^{\mu} u_c^{\nu}\,,
\end{equation}
where $e^{abc}$ is the covariant Levi-Civita symbol, and $s$ is a constant.
This leads us to restrict our considerations to backgrounds with totally
antisymmetric torsion. Indeed, axial spin tensor exclusively couples to the
axial component of torsion. Thus, without loss of generality, we shall
assume that $K^{\mu\nu\rho}$ is {\it totally antisymmetric}.

Let us now see how the membrane spin tensor of the form (\ref{jna8})
influences the equations (\ref{jna6}) and (\ref{jna7}). The computations are
straightforward, and have already been done in \cite{12b}. They lead to the
world-sheet equations
\begin{subequations} \label{jna9}
\begin{equation} \label{jna9a}
\nabla_a \left( m^{ab} u_{b\mu} \right) = \frac{s}{2} u^{\nu\lambda\rho}
K_{\mu\nu\lambda\rho}
\end{equation}
and boundary conditions
\begin{equation} \label{jna9b}
n_a \left( m^{ab} u_{b\mu} + \frac{3s}{2} e^{abc} K_{bc\mu} \right)
\Big|_{\del\cM} =0 \,.
\end{equation}
\end{subequations}
Here, the antisymmetric tensor $K_{\mu\nu\lambda\rho}$ is defined as
$$
K_{\mu\nu\lambda\rho} \equiv \del_{\mu} K_{\nu\lambda\rho} - \del_{\nu}
K_{\lambda\rho\mu} + \del_{\lambda} K_{\rho\mu\nu} - \del_{\rho}
K_{\mu\nu\lambda}\,,
$$
while $u^{\mu\nu\rho} \equiv e^{abc} u_a^{\mu} u_b^{\nu} u_c^{\rho}$, and
$K_{ab\rho}\equiv u^{\mu}_a u^{\nu}_b K_{\mu\nu\rho}$ are introduced for
convenience. The residual free coefficients $m^{ab}$ are related to the
original $t^{ab}$ as $m^{ab}\equiv t^{ab}-\frac{s}{2}\gamma^{ab}
u^{\mu\nu\rho} K_{\mu\nu\rho}$.

The world-sheet equations (\ref{jna9a}), and boundary conditions
(\ref{jna9b}) describe a membrane with maximally symmetric distribution of
spin, and arbitrary distribution of stress-energy. In \cite{12b}, the
maximal symmetry has been attributed to the membrane stress-energy, too.
This case is defined by the choice $m^{ab} = T \gamma^{ab}$, where $T$ is a
constant commonly interpreted as the membrane tension. With this, the
world-sheet equations and boundary conditions have been shown to follow from
an action functional. The resemblance of this action to the $\sigma$-model
action of \cite{22} has been pointed out in \cite{12b}. The two actions
differ in one instance only: the role of the 3-form field $B_{\mu\nu\rho}$
in \cite{22} is played by the contortion field $\frac{s}{T}K_{\mu\nu\rho}$
in \cite{12b}.

In the next section, we shall demonstrate how effective string dynamics is
obtained in the narrow membrane limit. The starting point of our analysis
will be the equations (\ref{jna9}). They defer from those used in \cite{12b}
by the form of $m^{ab}$ coefficients, which are allowed to violate the
condition $m^{ab}=T\gamma^{ab}$. As a consequence, we shall be able to
derive additional string couplings to the background geometry.

\section{Dimensional reduction}\label{Sec4}

The results of the preceding section are obtained under very general
assumptions concerning the dimensionality and topology of spacetime and
world-sheet. In what follows, we shall use this freedom to apply these
results to a cylindrical membrane wrapped around the extra compact dimension
of a $(D+1)$-dimensional spacetime. In the limit of a narrow membrane, we
expect to obtain the effective string dynamics. This kind of double
dimensional reduction has already been considered in \cite{22}. There, the
string effective action in $10$ dimensions has been obtained from the
membrane action in $11$ dimensions. In what follows, a similar $D+1\to D$
dimensional reduction will be applied to classical membranes in
Riemann-Cartan backgrounds.

Let us consider a $(D+1)$-dimensional spacetime with one small compact
dimension. It is parametrized by the coordinates $X^M$ ($M=0,1,\dots,D$),
which we divide into the ``observable'' coordinates $x^{\mu}$ ($\mu =
0,1,\dots,D-1$), and the extra periodic coordinate $y$. In the limit of
small extra dimension, we use the Kaluca-Klein ansatz
\begin{equation} \label{jna11}
\del_y K_{MNL} =0, \qquad \del_y G_{MN}=0
\end{equation}
to model the contortion and metric. This ansatz violates the
$(D+1)$-dimensional diffeomorphisms, leaving us with the coordinate
transformations
\begin{equation} \label{jna12}
x^{\mu'} = x^{\mu'} (x), \qquad y' = y + \lc(x).
\end{equation}
In what follows, we shall use the decomposition
\begin{equation} \label{jna13}
G_{MN} = \left(
\begin{array}{cc}
g_{\mu\nu} + \phi\,a_{\mu}a_{\nu} & \phi\,a_{\mu} \\ \phi\,a_{\nu} & \phi
\\
\end{array} \right) ,
\end{equation}
as it yields the variables that transform as tensors with respect to the
residual D-diffeomorphisms. The same kind of argument applies to $K_{MNL}$.
We shall use a \textit{totally antisymmetric contortion}, and decompose it
as
$$
K_{MNL} = \Big( K_{\mu\nu\lambda}\,,\, K_{\mu\nu y} \Big) ,
$$
with
\begin{equation} \label{jna14}
\begin{array}{rcl}
K_{\mu\nu y} & \equiv & k_{\mu\nu} \,,                     \\
K_{\mu\nu\lambda} & \equiv & \cK_{\mu\nu\lambda} + k_{\mu\nu}a_{\lambda} +
k_{\nu\lambda}a_{\mu} + k_{\lambda\mu}a_{\nu} \,.
\end{array}
\end{equation}
With respect to the complete residual transformations (\ref{jna12}), the
variables $\cK_{\mu\nu\lambda}$ and $k_{\mu\nu}$, as well as $g_{\mu\nu}$
and $\phi$, transform as tensors, while $a_{\mu}$ transforms as a
connection: $a_{\mu}^{\prime} = \left(a_{\nu}-\del_{\nu}\lc\right)\del
x^{\nu}/\del x^{\prime\mu}$.

Now, we consider a membrane wrapped around the extra compact dimension $y$.
Its world-sheet $X^M=Z^M(\xi^A)$ is denoted by $\cM_3$, and is chosen in the
form
\begin{equation} \label{jna15}
x^{\mu} = z^{\mu}(\xi^a), \qquad y=\xi^2,
\end{equation}
where the world-sheet coordinates $\xi^A$ ($A=0,1,2$) are divided into
$\xi^a$ ($a=0,1$) and $\xi^2$. This ansatz reduces the reparametrizations
$\xi^{A'}= \xi^{A'}(\xi^B)$ to
\begin{equation} \label{jna16}
\xi^{a'} = \xi^{a'}(\xi^b), \qquad \xi^{2'} = \xi^2 + \lc(z^{\mu}(\xi)),
\end{equation}
and the world-sheet tangent vectors $U_A^M = \del Z^M / \del\xi^A$ to
$$
U_a^{\mu} = u_a^{\mu}, \qquad U_2^{\mu} = U_a^y =0, \qquad U_2^y=1.
$$
One can verify that $u_a^{\mu} \equiv \del z^{\mu} / \del\xi^a$ transforms
as a tensor with respect to the residual spacetime and world-sheet
diffeomorphisms. The induced metric $\itGamma_{AB} \equiv G_{MN} U_A^M
U_B^N$ is shown to satisfy the condition $\del_2 \itGamma_{AB} =0$. It is
decomposed as
$$
\itGamma_{AB} = \left(
\begin{array}{cc}
\gamma_{ab} + \phi\,a_aa_b & \phi\,a_a \\
\phi\,a_b                  & \phi
\end{array} \right) ,
$$
with $\gamma_{ab}\equiv g_{\mu\nu} u_a^{\mu} u_b^{\nu}$, and $a_a \equiv
a_{\mu} u_a^{\mu}$. In what follows, we shall refer to
$x^{\mu}=z^{\mu}(\xi^a)$ as the string world-sheet, and denote it by
$\cM_2$.

The membrane boundary $\del\cM_3$ is given by $\xi^A = \zeta^A(\lambda^i)$,
where $\lambda^i$ ($i=0,1$) are the boundary coordinates. In accordance with
the ansatz (\ref{jna15}), it is chosen in the form
\begin{equation} \label{jna19}
\xi^a = \zeta^a(\lambda^0), \qquad \xi^2 = \lambda^1 .
\end{equation}
The boundary tangent vectors $V^A_i \equiv \del \zeta^A / \del\lambda^i$ are
then reduced to
$$
V^a_0 = v^a , \qquad V^a_1 = V^2_0 =0 , \qquad V^2_1 = 1 \,,
$$
and the boundary metric $\delta_{ij}\equiv\itGamma_{AB}V^A_iV^B_j$ becomes
$$
\delta_{ij} = \left(
\begin{array}{cc}
v^2 + \phi\,a^2 & \phi\,a \\
\phi\,a         & \phi
\end{array} \right) .
$$
Here, $v^a\equiv d\zeta^a/d\lambda^0$, $v^2 \equiv \gamma_{ab}v^av^b$ and $a
\equiv v^aa_a$. The boundary normal $N_A\equiv e_{ABC}V_0^B V_1^C/
\sqrt{-\delta}$ then reduces to $N_A=(n_a\,,0)$, with $n_a\equiv e_{ab}v^b/
\sqrt{-v^2}$. In what follows, we shall refer to $\xi^a =
\zeta^a(\lambda^0)$ as the string boundary $\del\cM_2$.

In this section, the dimensional reduction is applied to the membrane
equations (\ref{jna9}) of section \ref{Sec3}. In $(D+1)$-dimensional
background, the word-sheet equations (\ref{jna9a}) are rewritten as
\begin{subequations} \label{jna21}
\begin{equation} \label{jna21a}
\nabla_A \left(m^{AB}U_{BM}\right) = \frac{s}{2}U^{NLR}K_{MNLR}\,,
\end{equation}
while boundary conditions (\ref{jna9b}) become
\begin{equation} \label{jna21b}
N_A\left(m^{AB}U_{BM}+\frac{3s}{2}e^{ABC}K_{BCM}\right)\Big|_{\del\cM_3}=0\,.
\end{equation}
\end{subequations}
The dimensional reduction of (\ref{jna21}) with $m^{AB}=T\itGamma^{AB}$ has
already been studied in \cite{12b}. Here, we shall consider more general
situation
$$
m^{AB} = T \itGamma^{AB} + \mu^{AB} ,
$$
with $\mu^{AB}\neq 0$, and $\del_2\mu^{AB}=0$, in accordance with the
adopted Kaluza-Klein ansatz. Notice, however, that leaving the stress-energy
$\mu^{AB}$ completely arbitrary produces an inhomogeneous effective string
after dimensional reduction. This is not what we want. As we explained in
the introduction, our idea is to violate the maximal symmetry of the
membrane stress-energy $m^{AB}=T\itGamma^{AB}$ in a way which will preserve
the maximal symmetry of the effective string after dimensional reduction. To
this end, we are led to retain $m^{ab}=T\gamma^{ab}$ while leaving the other
$m^{AB}$ components unconstrained. In terms of $\mu^{AB}$, this means that
we must adopt $\mu^{ab}=0$. In what follows, we shall use the decomposition
\begin{equation} \label{jna23}
\mu^{AB} = \left(
\begin{array}{cc}
  0   &      \jmath^a       \\
 \jmath^b  & \omega-2\jmath^ca_c
\end{array} \right) ,
\end{equation}
as it yields parameters that transform as tensors with respect to the
residual reparametrizations. Indeed, the remaining free coefficients
$\jmath^a(\xi)$ and $\omega(\xi)$ are shown to transform covariantly under
(\ref{jna16}). We shall see later that $\jmath^a$ and $\omega$ are related
to the electric and dilatonic charges of the effective string. While the
condition $\mu^{ab}=0$ ensures maximal symmetry of the string stress-energy,
it does not ensure uniform distribution of its electric and dilatonic
charges. Nevertheless, in what follows, we shall keep the currents
$\jmath^a$ and $\omega$ arbitrary.

Let us now solve the world-sheet equations (\ref{jna21a}). First, we
decompose (\ref{jna21a}) into components parallel and orthogonal to the
world-sheet. The parallel component yields the conservation equation
$\nabla_A m^{AB}=0$. Using the Kaluca-Klein ansatz (\ref{jna11}),
(\ref{jna15}), (\ref{jna19}), and the decompositions (\ref{jna13}),
(\ref{jna14}) and (\ref{jna23}), it is reduced to
\begin{subequations} \label{jna24}
\begin{equation}\label{jna24a}
2\phi f_{ab}\jmath^b - \omega\del_a\phi = 0 \,,
\end{equation}
\begin{equation}\label{jna24b}
2\phi\nabla_a \jmath^a + 3\jmath^a\del_a\phi = 0 \,,
\end{equation}
with $f_{ab}\equiv\del_b a_a -\del_a a_b$. The $M=y$ component of the
remaining orthogonal equations is then shown to be identically satisfied,
while $M=\mu$ components become
\begin{eqnarray}
&& 2T\Big(\nabla_a u^a_{\mu}-\frac{1}{2\phi} P^{\nu}_{\orto\mu}
      \del_{\nu}\phi\Big) - \frac{3s}{\sqrt{\phi}}
      u^{\nu\lambda} k_{\mu\nu\lambda}          \nonumber  \\
&& +2\phi f_{\mu\nu}\jmath^{\nu}-\omega\del_{\mu}\phi = 0\,.  \label{jna24c}
\end{eqnarray}
\end{subequations}
The totally antisymmetric tensor $k_{\mu\nu\lambda}$ is defined as
$$
k_{\mu\nu\lambda} \equiv \del_{\mu} k_{\nu\lambda} + \del_{\nu}
k_{\lambda\mu} + \del_{\lambda} k_{\mu\nu} \,,
$$
while $f_{\mu\nu} \equiv \del_{\nu}a_{\mu} - \del_{\mu}a_{\nu}$, and
$\jmath^{\mu} \equiv u^{\mu}_a \jmath^a$.

The world-sheet equations (\ref{jna24}) show how effective string couples to
the background metric, torsion, electromagnetic and scalar fields. These
equations can be simplified by removing (\ref{jna24a}), as it coincides with
the parallel component of (\ref{jna24c}). Also, we are free to make
redefinitions of the background fields and free parameters. As it turns out,
the world-sheet equations (\ref{jna24}) are nicely simplified when expressed
in terms of the rescaled metric
$$
\tilde g_{\mu\nu} \equiv \sqrt{\phi}\, g_{\mu\nu} \,,
$$
and rescaled current
$$
\tilde\jmath^a \equiv \phi \jmath^a \,.
$$
Indeed, in terms of $\tilde g_{\mu\nu}$ and $\tilde \jmath^a$, the string
dynamics (\ref{jna24}) is rewritten as
\begin{subequations} \label{jna26}
\begin{equation}\label{jna26a}
\tilde\nabla_a \tilde \jmath^a = 0 \,,
\end{equation}
\begin{equation}\label{jna26b}
T\tilde\nabla_a u^a_{\mu}=\frac{3s}{2}
      \tilde u^{\nu\lambda}k_{\mu\nu\lambda}
      - f_{\mu\nu}\tilde \jmath^{\nu} + \frac{\omega}{2}
      \del_{\mu}\phi \,.
\end{equation}
\end{subequations}
The world-sheet equations (\ref{jna26}) govern the dynamics of the effective
string which carries more charges than the mere tension and spin. As a
consequence, the string is coupled not only to the metric and torsion, as
described in \cite{12b}, but also to the electromagnetic and scalar fields.

The dimensional reduction of boundary conditions (\ref{jna21b}) is performed
in a similar way. In terms of $\tilde g_{\mu\nu}$ and $\tilde \jmath^a$, the
resulting equations take the form
\begin{subequations} \label{jna27}
\begin{equation}\label{jna27a}
\tilde n_a \tilde \jmath^a \big|_{\del\cM_2} = 0 \,,
\end{equation}
\begin{equation}\label{jna27b}
\tilde n_a\left(T u^a_{\mu}+3s\,\tilde e^{ab} u^{\nu}_b
      k_{\mu\nu}\right)\big|_{\del\cM_2} = 0 \,.
\end{equation}
\end{subequations}
As we can see, the electromagnetic field $a_{\mu}$ does not appear in the
boundary conditions. This may lead us to the conclusion that the behavior of
our effective strings in Riemann-Cartan spacetime is quite different from
that of fundamental strings in the low-energy string backgrounds. In the
next section, we shall compare our effective string equations with the
string $\sigma$-model of \cite{15,16,17,18,t7,t19}, and demonstrate that it
is not quite so.

As a preparation for this comparison, let us transform our equations
(\ref{jna26}), (\ref{jna27}) to a more convenient form. We start with the
observation that (\ref{jna26b}) implies the constraint $\omega \tilde
\jmath^{\lambda} \del_{\lambda} \phi = 0$, which reduces to
$$
\tilde \jmath^a \del_a \phi = 0
$$
in the generic case $\omega \neq 0$. The general solution of this constraint
has the form
$$
\tilde \jmath^a = e\, e^{ab} \del_b \phi \,
$$
where $e(\xi)$ is a residual coefficient that defines the distribution of
electric charge along the string. In what follows, we shall adopt
$$
e = {\rm const.}
$$
in accordance with our systematic consideration of strings with maximally
symmetric distribution of matter. With this, the conservation equation
(\ref{jna26a}) is identically satisfied, while the world-sheet equations
(\ref{jna26b}) become
$$
T\tilde\nabla_a u^a_{\mu} = \frac{3s}{2} \tilde u^{\nu\lambda} \left(
k_{\mu\nu\lambda} - \frac{2e}{3s} f_{\mu\nu} \del_{\lambda}\phi \right)
+\frac{\omega}{2} \del_{\mu}\phi \,.
$$
Now, we are free to make additional redefinitions of the background fields
and free parameters. As it turns out, the world-sheet equations are nicely
simplified when expressed in terms of the redefined torsion
$$
\tilde k_{\mu\nu} \equiv k_{\mu\nu} - \frac{e}{3s} \phi f_{\mu\nu} \,,
$$
and redefined dilatonic charge
$$
\tilde\omega \equiv \omega + e u^{\mu\nu} f_{\mu\nu} \,.
$$
Indeed, in terms of $\tilde k_{\mu\nu}$ and $\tilde\omega$, our world-sheet
equations are rewritten as
\begin{equation}\label{jna28}
T\tilde\nabla_a u^a_{\mu} = \frac{3s}{2} \tilde u^{\nu\lambda} \tilde
k_{\mu\nu\lambda} +\frac{\tilde\omega}{2} \del_{\mu}\phi \,.
\end{equation}
As we can see, the resultant world-sheet equations contain no coupling to
the electromagnetic field.

Let us now see how the above results influence the boundary conditions
(\ref{jna27}). First, we notice that (\ref{jna28}) implies the constraint
$\tilde\omega\,u^{\mu}_a \del_{\mu} \phi = 0$, which reduces to
$\del_a\phi=0$ in the generic case $\tilde\omega\neq 0$. This means that our
effective string is forced to live on the surface $\phi={\rm const}$. As a
consequence, the boundary condition (\ref{jna27a}), which reduces to $v^a
\del_a \phi = 0$, is identically satisfied. The remaining boundary
conditions (\ref{jna27b}) are then rewritten in terms of $\tilde k_{\mu\nu}$
and $\tilde e\equiv e\phi$ as
\begin{equation}\label{jna29}
\tilde n_a \left[T u^a_{\mu} + \tilde e^{ab} u^{\nu}_b \left( 3s\,\tilde
k_{\mu\nu} + \tilde e f_{\mu\nu} \right) \right]\big|_{\del\cM_2} = 0 \,.
\end{equation}
As we can see, the coupling to the electromagnetic field reappears in the
boundary conditions.

To summarize, we have shown that the effective string dynamics is governed
by the world-sheet equations (\ref{jna28}), and boundary conditions
(\ref{jna29}). The equations are parametrized by four parameters, $T$, $s$,
$\tilde e$, and $\tilde\omega$, which define the string tension, spin,
electric charge and dilatonic charge, respectively. The constant parameters
$T$, $s$ and $\tilde e$ define uniform distribution of stress-energy, spin
and electric charge, while dilatonic charge $\tilde\omega$ is left
arbitrary. In the next section, this form of string dynamics will be shown
to follow from an action functional that coincides with the string theory
$\sigma$-model discussed in \cite{15,16,17,18,t7,t19}.

\section{Comparison with the string sigma model}\label{Sec5}

In this section, we shall compare our results with the predictions of the
string $\sigma$-model \cite{15,16,17,18,t7,t19}. This model is defined by
the action functional
\begin{eqnarray}
I = &&  T\int d^2\xi \sqrt{-h} \Big[
        G_{\mu\nu}(x) u_a^{\mu} u_b^{\nu} h^{ab} \nonumber \\
&& +B_{\mu\nu}(x)u_a^{\mu}u_b^{\nu}e^{ab}+\Phi(x)R^{(2)}\Big], \label{jna30}
\end{eqnarray}
in which $x^{\mu}(\xi)$ and $h_{ab}(\xi)$ are considered independent
variables. The string background fields $G_{\mu\nu}(x)$, $B_{\mu\nu}(x)$ and
$\Phi(x)$ are commonly referred to as metric, string axion and dilaton,
respectively. The $2d$ curvature $R^{(2)}$ is constructed out of the
auxiliary metric $h_{ab}$.

By varying the action (\ref{jna30}) with respect to $x^{\mu}$ one finds the
world-sheet equations
\begin{subequations} \label{jna31}
\begin{equation} \label{jna31a}
\nabla_a u^a_{\mu} = \frac{1}{2} u^{\nu\lambda}B_{\mu\nu\lambda} +
\frac{1}{2} R^{(2)} \del_{\mu}\Phi \,,
\end{equation}
and boundary conditions
\begin{equation} \label{jna31b}
n_a\left( u^a_{\mu} + e^{ab} u_b^{\nu} B_{\mu\nu}\right)\big|_{\del\cM_2} =
0 \,,
\end{equation}
with $\nabla_a = \nabla_a (h)$, $R^{(2)} = R^{(2)}(h)$, and
$B_{\mu\nu\lambda} \equiv \del_{\mu} B_{\nu\lambda} + \del_{\nu}
B_{\lambda\mu} + \del_{\lambda} B_{\mu\nu}$. The variation with respect to
$h^{ab}$, on the other hand, gives
\begin{equation} \label{jna31c}
\gamma_{ab} - \frac{1}{2} \left(\gamma_{cd}h^{cd}\right) h_{ab} =
\nabla_a\nabla_b \Phi - h_{ab} \nabla^2 \Phi \,,
\end{equation}
\end{subequations}
with $\nabla^2 \equiv h^{ab}\nabla_a\nabla_b$. These equations are nicely
simplified by noticing that (\ref{jna31a}) implies the constraint $R^{(2)}
u^{\mu}_a \del_{\mu}\Phi = 0$. In the generic situation, characterized by
the non-vanishing scalar curvature $R^{(2)}$, it takes the simple form
$$
\del_a \Phi = 0 \,,
$$
telling us that the string is forced to live on the surface $\Phi={\rm
const}$. The equations (\ref{jna31c}) are thereby reduced to
\begin{equation} \label{jna33}
\gamma_{ab} = e^{\chi} h_{ab} \,,
\end{equation}
where $\chi(\xi)$ is an arbitrary function on the world-sheet. The equation
(\ref{jna33}) can be used to replace the auxiliary variable $h_{ab}$ with
$\gamma_{ab}$ in all but the last term of (\ref{jna31a}). Indeed, in terms
of $\gamma_{ab}$, the world-sheet equations (\ref{jna31a}) are rewritten as
$$
\nabla_a u^a_{\mu} = \frac{1}{2} u^{\nu\lambda}B_{\mu\nu\lambda} +
\frac{1}{2}\left( R^{(2)} - \nabla^2\,\chi \right)\del_{\mu}\Phi \,,
$$
while boundary conditions (\ref{jna31b}) are shown to retain the same form.
This time, however, $\nabla_a = \nabla_a(\gamma)$, $R^{(2)} =
R^{(2)}(\gamma)$ etc. As we can see, the conformal factor $\chi$ of the
auxiliary metric $h_{ab}$ remains in the field equations as a free
coefficient. As $\chi(\xi)$ is an unconstrained arbitrary function, we are
free to make the redefinition
$$
\Omega \equiv R^{(2)} - \nabla^2\,\chi \,,
$$
thereby bringing the field equations to the simple form
\begin{subequations} \label{jna34}
\begin{equation} \label{jna34a}
\nabla_a u^a_{\mu} = \frac{1}{2} u^{\nu\lambda}B_{\mu\nu\lambda} +
\frac{\Omega}{2}\,\del_{\mu}\Phi \,,
\end{equation}
\begin{equation} \label{jna34b}
n_a\left( u^a_{\mu} + e^{ab} u_b^{\nu} B_{\mu\nu}\right)\big|_{\del\cM_2} =
0 \,.
\end{equation}
\end{subequations}
The new free coefficient $\Omega(\xi)$ is a remnant of the auxiliary
variable $h_{ab}$.

The string dynamics in the form (\ref{jna34}) lacks the interaction with the
background electromagnetic field $A_{\mu}$. This interaction is
conventionally introduced by the replacement
$$
B_{\mu\nu} \to B_{\mu\nu} + F_{\mu\nu}\,,
$$
where $F_{\mu\nu}\equiv \del_{\nu} A_{\mu} - \del_{\mu} A_{\nu}$.
Being the gauge transform of the $B_{\mu\nu}$ field, this replacement does
not influence the world-sheet equations, as they depend on $B_{\mu\nu}$ only
through the gauge invariant field strength $B_{\mu\nu\lambda}$. On the other
hand, the boundary conditions (\ref{jna34b}) are not gauge invariant. The
replacement $B \to B + F$ brings them to the form
\begin{equation} \label{jna35}
n_a\left[ u^a_{\mu} + e^{ab} u_b^{\nu}\left( B_{\mu\nu} + F_{\mu\nu}
\right) \right] \big|_{\del\cM_2} = 0
\end{equation}
showing that the string ends have nontrivial coupling to the electromagnetic
field.

Now, we are ready to compare our equations (\ref{jna28}), (\ref{jna29}) with
the string $\sigma$-model equations (\ref{jna34}), (\ref{jna35}). The first
set of equations describes an effective string obtained by dimensional
reduction of a narrow membrane in $(D+1)$-dimensional Riemann-Cartan
spacetime. The second set governs the dynamics of an elementary string
coupled to the string background fields $G_{\mu\nu}$, $B_{\mu\nu}$,
$A_{\mu}$ and $\Phi$. By comparing the two sets of equations, we find that
they differ in one instance only: the role of the string background fields
in (\ref{jna34}), (\ref{jna35}) is played by the dimensionally reduced
Riemann-Cartan geometry in (\ref{jna28}), (\ref{jna29}). Precisely, the
identification of external fields
\begin{eqnarray}
&& \tilde g_{\mu\nu}(x) \to G_{\mu\nu}(x)\,, \quad
   3s\,\tilde k_{\mu\nu}(x) \to T B_{\mu\nu}(x)\,,      \nonumber \\
&& \tilde e\, a_{\mu}(x) \to T A_{\mu}(x)\,, \quad
   \phi(x) \to T \Phi(x) \,,                            \nonumber
\end{eqnarray}
and free coefficients
$$
\tilde\omega(\xi) \to \Omega(\xi)
$$
establishes $1-1$ correspondence between the two theories. Whether this is
just a coincidence, or there is more content in this matching is not the
subject of this work. Anyhow, we find it interesting enough to justify our
presentation.

In summary, we have derived how effective string, obtained by dimensional
reduction of a narrow membrane, behaves in a dimensionally reduced
Riemann-Cartan spacetime. We have considered an effective string made of
uniformly distributed spinning matter, thus representing a simple
generalization of the Nambu-Goto case. As a consequence, novel couplings to
the background geometry have been discovered. In particular, the effective
string has been shown to couple to the dimensionally reduced Riemann-Cartan
geometry the same way as fundamental string couples to the low-energy string
fields.

\section{Concluding remarks}\label{Sec6}

In this paper, we have analyzed the behavior of a narrow membrane wrapped
around the extra compact dimension of a $(D+1)$-dimensional Riemann-Cartan
spacetime. The membrane constituent matter is specified in terms of its
stress-energy and spin tensors. A membrane with maximally symmetric
distribution of stress-energy and spin has already been considered in
\cite{12b}. After dimensional reduction, such a membrane has been shown to
reduce to a string that couples to the metric and torsion the same way as
fundamental string couples to the low-energy string fields $G_{\mu\nu}$ and
$B_{\mu\nu}$. In this paper, we have relaxed the condition of maximal
symmetry used in \cite{12b} by allowing the membrane stress-energy to have
non-uniform distribution in the compact dimension. This way, the effective
string retains maximal symmetry, but is characterized by more free
parameters than the mere tension and spin. We have demonstrated that such
strings carry electric and dilatonic charges, and couple to the
electromagnetic and scalar fields the same way as fundamental strings couple
to the low-energy string fields $A_{\mu}$ and $\Phi$. In fact, {\it we have
established a $1-1$ correspondence between the macroscopic string dynamics
in Riemann-Cartan spacetime, and the fundamental string dynamics in the
low-energy string backgrounds}.

We have started our exposition in section \ref{Sec2} by reviewing the basics
of the multipole formalism developed in \cite{11,12,12a}. This method is a
generalization of the Mathisson-Papapetrou method for pointlike matter
\cite{1,2}. It has already been used in \cite{11,12,12a,12b} for the study
of strings and higher branes in Riemann-Cartan backgrounds. Its essence is
the usage of stress-energy and spin tensor conservation laws to specify the
brane dynamics. The advantage of this method is that it is model independent
as no action functional is specified. In section \ref{Sec3}, the basic
results of \cite{12b} have been reviewed. The single-pole solution of the
conservation equations has been applied to the membrane with maximally
symmetric distribution of spin. The obtained equations have been marked as a
starting point for the main analysis of the paper. This analysis has been
done in section \ref{Sec4}. We have considered cylindrical membranes wrapped
around the extra compact dimension of a $(D+1)$-dimensional spacetime. The
effective string dynamics has been obtained in the narrow membrane limit. As
compared to \cite{12b}, our effective string is characterized by two more
free parameters, and additionally couples to the electromagnetic and scalar
fields. In section \ref{Sec5}, we have compared our effective string
dynamics with the string theory $\sigma$-model of \cite{15,16,17,18,t7,t19}.
We have demonstrated that our macroscopic strings couple to the
dimensionally reduced Riemann-Cartan geometry the same way as fundamental
strings couple to the low-energy string fields.

In conclusion, let us say something about the prospects of our research. We
have already established a $1-1$ correspondence between the macroscopic
string dynamics in Riemann-Cartan spacetime, and the fundamental string
dynamics in the low-energy string backgrounds. In this correspondence, the
string background fields $G_{\mu\nu}$, $B_{\mu\nu}$, $A_{\mu}$ and $\Phi$
are related to the metric and torsion of the Riemann-Cartan spacetime. An
interesting challenge would be to establish equivalence on the level of
background field equations. There have been attempts in literature to
rewrite the low energy string field action in geometric terms \cite{25,26}.
These have not been very successful though, as they included some unnatural
constraints to be imposed on torsion prior to varying the action. Note that
establishing equivalence on the level of background field equations is not
about comparing the existing equations, but rather about constructing the
gravity equations that match those of the low energy string fields. This is
because there is no a preferred geometric action with dynamical torsion that
could readily be compared with the low energy string field action. Indeed,
the Einstein-Hilbert action is torsion free, while the torsion of
Einstein-Cartan theory does not propagate. If we stay with membranes,
however, the construction of the needed action is quite simple. One should
start with the low energy string field action, and replace the symmetric
field with the spacetime metric, and the 3-form field with the axial
component of the torsion. It is much more complicated to obtain geometric
counterparts of the 2-form field, electromagnetic field and dilaton field.
The natural idea is to follow the dimensional reduction procedure used in
this paper. One should try and find a $(D+1)$-dimensional geometric action
that reduces to the needed string background action after dimensional
reduction. This is, however, a difficult task for itself and should be
considered in a separate paper.

\acknowledgments
This work is supported by the Serbian Ministry of Science and Technological
Development, under Contract No. $141036$.

\end{document}